\documentclass[%
twocolumn,
amsmath,amssymb,
aps,
prl, 10pt
]{revtex4-2}

\usepackage{graphicx}
\usepackage{dcolumn}
\usepackage{bm}
\usepackage{subcaption}

\usepackage{hyperref}

\hypersetup{
	colorlinks=true,
	linkcolor=blue,
	urlcolor=magenta,
	citecolor=red,
}

\usepackage{float}
\allowdisplaybreaks
\def\be{\begin{equation}}
	\def\ee{\end{equation}}
\def\half{\frac{1}{2}}
\begin{document}
	
	\preprint{APS/123-QED}
	
	\title{Ramanujan's $1/\pi$ series and conformal field theories}
	
	\author{Faizan Bhat$^{a}$}
	\email{faizanbhat@iisc.ac.in}
	\author{Aninda Sinha$^{a,b}$}%
	\email{asinha@iisc.ac.in}
	\affiliation{%
		${^a}$Centre for High Energy Physics, 
		Indian Institute of Science, 
		C.V. Raman Avenue, Bangalore 560012, India.\\
        \it ${^b}$Department of Physics and Astronomy, University of Calgary, \it Alberta T2N 1N4, Canada.
	}%
	
	\date{\today}
	
\begin{abstract}

In 1914, Ramanujan unveiled 17 extraordinary infinite series for $1/\pi$. In this work, we uncover their physics origin by relating them to 2D logarithmic conformal field theories (LCFTs), which emerge in diverse settings such as the fractional quantum Hall effect, percolation, polymers, and even holography. Through this LCFT connection, we reinterpret such infinite series in terms of fundamental CFT data — the operator spectrum and OPE coefficients. This perspective leads to novel physics-inspired approximations for
$1/\pi$. Drawing lessons from Ramanujan's formulae, we construct a new family of bases for expanding LCFT correlators that converge far more rapidly than the standard conformal block decomposition. This is achieved using recently developed stringy/parametric crossing-symmetric dispersion relations. Remarkably, when working with these new expansions, the action of a certain differential operator (which arises naturally from the Ramanujan connection) dramatically enhances convergence, with the entire contribution collapsing to that of the logarithmic identity operator. This striking simplification hints at a universal property of LCFTs. Finally, we discuss a new holographic interpretation of this unexpected mathematics–physics connection.

\end{abstract}

\maketitle	
\textbf{Introduction:}
In 1914, Ramanujan \cite{ram} recorded 17 remarkable formulae for $\pi$ of the form
\begin{equation}\label{ram}
\sum_{n=0}^\infty \frac{(\frac{1}{2})_n (\sigma)_n(1-\sigma)_n}{n!^3} (a+b n) z_0^n=\frac{1}{\pi}\,.
\end{equation}
Here, $\sigma \in \{\frac{1}{2},\frac{1}{3},\frac{1}{4},\frac{1}{6}\}$, while $z_0, b,$ and $a$ are algebraic numbers. At present, many such Ramanujan-like formulae are known \cite{pirev}.  These formulae not only laid the foundation for all subsequent fast-converging series for $\pi$ but also led to several breakthroughs that continue to shape mathematics even today, especially analytic number theory and computational mathematics. In this work, we show that these formulae have a natural physics interpretation in terms of four-point correlators in 2D logarithmic conformal field theories (LCFTs) \cite{saleur, gurarie, flohr}. LCFTs are a fascinating class of non-unitary CFTs. They describe statistical physics systems such as critical dense polymers \cite{saleur}, percolation \cite{cardy, flohrlohmann, langlands}, and the fractional quantum Hall effect \cite{nayak}. They also appear in the study of celestial CFTs \cite{bissiCCFT} and in the context of holography, describing the dynamics of a scalar field of a particular mass in the AdS Schwarzschild black brane background \cite{grinberg}---this last connection is expanded on in the Appendix. The fact that Ramanujan's extraordinary formulae naturally arise in LCFTs opens up avenues for further exploration at the intersection of physics and mathematics.

All Ramanujan-like formulae originate from the Legendre relation, which is given as \cite{guillera, wolframram}
\begin{equation}\label{leg}
z(z-1) \left(F_\sigma(z)\partial_{z}F_\sigma(1-z)-F_\sigma(1-z)\partial_{z} F_\sigma(z)\right)=\frac{\sin \pi\sigma}{\pi}\,,   
\end{equation}
where
\begin{equation}
\label{hypdef}
F_\sigma(z)={}_2F_1(\sigma,1-\sigma,1,z) \,.
\end{equation}
An important clue to arrive at the physics connection lies in the analytic structure of $F_\sigma(z)$, which exhibits a logarithmic singularity at $z =1$. This points us toward 2D LCFTs, which are known for having logarithmic behaviour in their correlation functions, rather than the typical power-law behaviour. More precisely, we consider the four-point correlator of twist operators in the $c=-2$ LCFT, which is arguably the most well-studied LCFT in the literature \cite{saleur, gurarie, gaberdiel1, gaberdiel2, gaberdiel3, curious} --- we give more details about the $c=-2$ LCFT in the Appendix and comment on other central charges in \cite{supp}. The correlator has been explicitly calculated and is given as \cite{saleur, gurarie}. 
\begin{equation}
\begin{split}
\label{LCFTcorr}
    G(z,\bar{z}) &= \kappa(z,\bar{z})( F_\sigma(z)F_\sigma(1-\bar{z})+ F_\sigma(\bar{z})F_\sigma(1-z))\,,\\&
     \equiv G_L(z,\bar{z})+ G_R(z,\bar{z})\,,
\end{split}
\end{equation}
where $\kappa(z,\bar{z}) = |z(1-z)|^{2\sigma(1-\sigma)}$. We have used $\sigma$ to denote the twist of the operators, their conformal dimension $\Delta =\sigma(\sigma-1)$, and $z$ is the 2D conformal cross-ratio. Now, recall that the Witt algebra generators are given as $\ell_n\equiv -z^{n+1}\partial_z,\bar\ell_n\equiv -\bar z^{n+1}\partial_{\bar z}$. Defining the operator $\mathcal{L}=-(\ell_0-\bar \ell_0)+(\ell_1-\bar \ell_1)$, which is a linear combination of rotation and special conformal transformation, the Legendre relation in the LCFT language is simply:
\begin{equation}\label{legcft}
    {\mathcal L}G_L(z,\bar z){\bigg |}_{\bar z=z}= \kappa(z,z)\frac{\sin(\pi\sigma)}{\pi}=-{\mathcal L}G_R(z,\bar z){\bigg |}_{\bar z=z}\,.
\end{equation}
Thus, we see that the $\sigma$ appearing in Ramanujan's formulae \eqref{ram} is mapped to the twists or, equivalently, to the conformal dimensions of the twist operators in the LCFT. In the Appendix, we discuss a physical interpretation of the Legendre relation via holography.

Having expressed the Legendre relation in terms of an LCFT correlator, we can explicitly specify the conformal data that goes into it by performing a conformal block decomposition. Doing so leads us to new approximations for $\pi$ of the form given in \eqref{LegCBDexp}. We have not encountered such approximations in the literature (at least they do not appear to be commonly reported --- see e.g.\cite{pirev}). While novel and interesting, these formulae do not converge fast. This is because the blocks \( g_{\Delta,\ell}(z) \) (see~\eqref{2DBlockDef}) form a good basis only near \( z, \bar{z} \sim 0 \), while the correlator contains both \( F_\sigma(z) \) and \( F_\sigma(1 - z) \). As a result, for any \( 0 \le z \le 1 \), convergence is slow. Ramanujan encountered the same issue: directly plugging the series expansion for \(F_\sigma(z)\) and \( F_\sigma(1-z)\) into the Legendre relation does not yield fast-converging series for \( 1/\pi \). To resolve this, he came up with ingenious manipulations that exploited special properties of elliptic integrals and modular equations, which we will review later. The key idea of his techniques is to rewrite the Legendre relation, so only $F_{\sigma}(z)$ appears, and convergence becomes fast for \( z \sim 0 \).   

Ramanujan’s result motivates the following question: \textit{Can one emulate his procedure to derive a new, rapidly converging basis for expanding LCFT correlators in terms of conformal data?} We answer this question in the affirmative. The key idea, motivated by lessons derived from the derivation of Ramnujan's formulae, is to employ the recently derived stringy dispersion relation \cite{stringydisp} to express the full LCFT correlator solely in terms of its discontinuity across the logarithmic branch cut. Unlike the full correlator, this discontinuity admits a rapidly converging conformal block expansion around \( z, \bar{z} = 0 \), yielding an efficient basis for reconstructing the entire correlator. Remarkably, in this new basis, the Legendre relation in \eqref{legcft} converges incredibly fast; in fact, \textit{it is reproduced entirely from the contribution of the identity operator alone}.  We emphasise that this striking simplification arises only after the application of the \( \mathcal{L} \) operator--- such a structure would remain hidden without casting the Legendre relation in LCFT language. This result does not depend on the details of the LCFT, but only on the fact that the correlators possess logarithmic singularities, hinting at a universal feature of LCFTs.

\begin{figure}[H]
	\centering
	\includegraphics[scale=0.5]{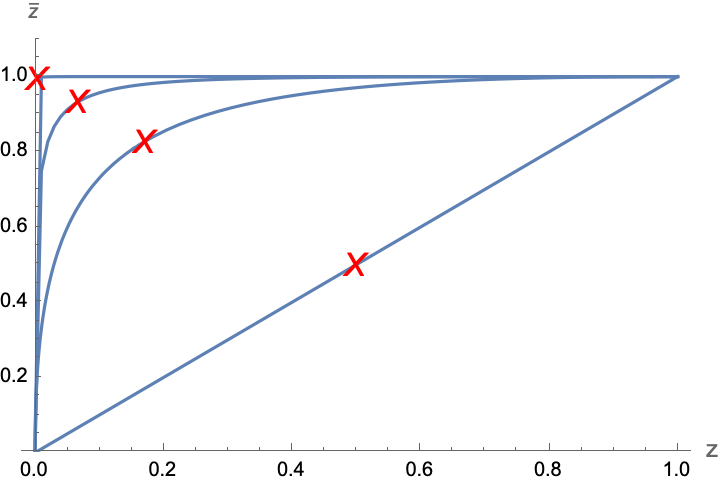}
	\caption{Slices in the $z,\bar{z}$ plane corresponding to solutions of the modular equations for $n=1,2,3,7$ from right to left. Red crosses indicate the singular values $z_0$.}\label{slice}
\end{figure}

{\bf The mathematics behind Ramanujan's $\mathbf{1/\pi}$ formulae:} Let us briefly review the derivation of Ramanujan's formulae, as it is very instructive. As mentioned before, directly plugging the series representations for $F_{\sigma}(z)$ and $F_{\sigma}(1-z)$ into the Legendre relation does not lead to fast-converging formulae for $1/\pi$. Any series that would arise from there would, first, be a double sum, and secondly, be very slowly convergent, since one sum needs $ z\sim 0$, and the other $z\sim 1$ for fast convergence. The first step, therefore, is to recast the Legendre relation so that only $F_\sigma(z)$ appears. This is achieved using the so-called modular equations. According to these, {for any rational $n$, whenever $z$ and $\bar{z}$ satisfy a "modular" condition $z=f_n(\bar z)$ \cite{borwein,supp}, the following holds}:
\begin{equation}\label{mod}
F_\sigma(1-z)F_\sigma(\bar z)= n F_\sigma(z)F_\sigma(1 -\bar z)\,,   
\end{equation}
Here, $n$ is called the degree of the modular equation, and we shall only consider $n \in \mathbb{N}$. The case $n=1$ is trivial, and the modular condition is just $z = \bar z$. As a non-trivial example, the modular condition for $n=3$ is
\begin{equation}\label{n3}
    (z\bar z)^{1/4}+[(1-z)(1-\bar z)]^{1/4}=1\,.
\end{equation}
Next, we solve the modular condition for singular values $z_0$ defined as $z_0 = f_n(1-z_0)$. From \eqref{mod}, we then get
\begin{equation}
              \sqrt{n}F_\half(z_0)=F_\half(1-z_0)\,.
\end{equation}
Plugging this into the Legendre relation, the LHS becomes
\begin{equation}
\frac{z_0(1-z_0)}{2} \left(\sqrt{n}\partial_{z} F^2_\sigma(z)|_{z=z_0}+\frac{1}{\sqrt{n}}\partial_{z}F^2_\sigma(z)|_{z=1-z_0}\right)\,.    
\end{equation}
The job is not done yet, as the two terms are still being evaluated at $z_0$ and $1-z_0$. Using \eqref{mod} again, after some lengthy algebra \cite{borwein}, we arrive at the final form:
\begin{equation}
\label{final}
    \sqrt{n}z_0(1-z_0)\left(\frac{d}{dz}-\frac{d M_n(\bar z)}{d\bar z}{\bigg |}_{\bar z=1-z_0}\right)F_\half(z)^2{\bigg |}_{z=z_0}=\frac{1}{\pi}
\end{equation}
where we have defined the multiplier $M_n(\bar z)$ as $M_n^2(\bar z)=\bar z(1-\bar z)/(z(1-z)) dz/d\bar z$ where $z = f_n(\bar{z})$. For $n=1$, $M_1(\bar z)=1$. For $n=3$, one can work out $dM_3/d\bar z|_{\bar z=1-z_0}=-4/\sqrt{3}$ with $z_0=(2-\sqrt{3})/4\approx 0.067$. At this point, we would get fast-converging formulae for $1/\pi$ by plugging the series representation for $F_{\sigma}(z)$, but it would still be a double sum. To get to the single sum that appears in Ramanujan's formulae, we use Clausen's identity given as (e.g. \cite{guillera})
\begin{equation}\label{clausen}
    F_\sigma(z)^2={}_3F_2(\sigma,1-\sigma,\frac{1}{2};1,1;4z(1-z))\,,
\end{equation}
for $z < 1/2$.
Plugging this into eq. \eqref{final} leads to Ramanujan-like formulae for all $n$. For example, for $n=3$, we get
\begin{equation}\label{rampi}
    \frac{4}{\pi}=\sum_{k=0}^\infty (6k+1) (\frac{1}{4})^k\frac{(\half)_k^3}{k!^3}\,.
\end{equation}
which is among the original 17 formulae Ramanujan presented. The fastest of his formulae arises from $n=58$. In general, convergence improves as $n$ increases because the singular values $z_0$ approach $0$ along the line $\bar{z}=1-z$. We show this in Fig. \ref{slice} using the results in \cite{borwein}.

The derivation of Ramanujan’s formulae offers valuable insights, with two key lessons. First, it demonstrates that the Legendre relation can be transformed---through a sequence of highly non-trivial manipulations---into a form involving only a power series expansion around \( z = 0 \), resulting in remarkably fast convergence. Second, the appearance of a parameter \( n \), corresponding to the order of the modular equations, introduces a family of such formulae. Each value of \( n \) yields a distinct series representation for \( 1/\pi \), with the rate of convergence improving significantly as \( n \) increases. In light of the LCFT connection, how do we leverage these lessons to learn more about the LCFT correlators? Or from a different perspective, can we give a physics explanation for the existence of a one-parameter family of fast-converging Ramanujan-like formulae? In the following sections, we address these questions. 

{\bf{The CFT Legendre Relation}}: In \eqref{legcft}, we recast the Legendre relation in terms of the four-point correlator of twist operators in the well-studied $c = -2$ LCFT. From the CFT perspective, the natural basis for expanding correlators is in terms of the (global) conformal blocks, labelled by the scaling dimensions and spins of the primary operators in the theory. As explained in \cite{abcd}, in the case of LCFTs,  the correct version of the conformal blocks also involves derivatives of the usual blocks, leading to the following "logarithmic" conformal block decomposition.
\begin{equation}
    G(z,\bar{z})=  (z \bar{z})^{\sigma(1-\sigma)} \sum_{\Delta,\ell} (c^{(1)}_{\Delta,\ell} + c^{(2)}_{\Delta,\ell}\partial_{\Delta})g_{\Delta,\ell}(z,\bar{z})
\end{equation}
$g_{\Delta,\ell}(z,\bar{z})$ are the 2D conformal blocks given as
\begin{equation}
\begin{split}
\label{2DBlockDef}
   g_{\Delta,\ell}(z,\bar{z}) &= k_{\Delta+\ell}(z) k_{\Delta-\ell}(\bar{z})+ k_{\Delta+\ell}(\bar{z}) k_{\Delta-\ell}(z), \\
   k_{\beta}(x) &= x^{\frac{\beta}{2}}{}_2F_1(\frac{\beta}{2},\frac{\beta}{2},\beta,x)\,.
\end{split}
\end{equation}
Our goal is to write down the conformal block expansion of the CFT Legendre relation. Note that the Legendre relation only involves $G_{L/R}(z,\bar{z})$, not the full correlator $G(z,\bar{z})$.  However, we can express $G_{L/R}(z,\bar{z})$ in terms of the logarithmic part of the full correlator $G_{log}(z,\bar{z})$ or, equivalently, in terms of the discontinuity of the correlator across the logarithmic branch cut, since $G_{disc}(z,\bar{z}) = \pi G_{log}(z,\bar{z})$.  We have
\begin{equation}
\begin{split}
\label{LogToLeft}
     G_{log}(z,\bar{z}) &=  -\frac{\sin(\pi \sigma)}{\pi}\kappa(z,\bar{z})F_{\sigma}(z) F_{\sigma}(\bar{z}) \\
          &=  -\frac{\sin(\pi \sigma)}{\pi}G_{L}(z,1-\bar{z})\,.
\end{split}
\end{equation}
Now, it is easy to see that the block expansion of the discontinuity only involves the $c^{(2)}_{\Delta,\ell}$ data. 
\begin{equation}
\label{logcorCBD}
\begin{split}
 G_{log}(z,\bar{z}) &=  (z \bar{z})^{\sigma(1-\sigma)}\sum_{\Delta,\ell} \frac{c^{(2)}_{\Delta,\ell}}{2} g_{\Delta,\ell}(z,\bar{z}) \,.
\end{split}
\end{equation}
We will refer to the operators linked to the $c^{(2)}_{\Delta,\ell}$-data as log-operators. Using $g_{\Delta,\ell}(z,\bar{z}) \underset{z,\bar{z} \to 0}{\sim}  z^{\frac{\Delta -\ell}{2}}\bar{z}^{\frac{\Delta+\ell}{2}}+z^{\frac{\Delta +\ell}{2}}\bar{z}^{\frac{\Delta-\ell}{2}}$, we match the powers on both sides to extract the spectrum and the OPE coefficients.  The spectrum is given as $(\Delta,\ell) = (\ell+2n, \ell)$, $\ell,n \in \mathbb{Z}^{\ge 0}$. All OPE coefficients $c^{(2)}_{\Delta,\ell}$ hide a factor $\frac{\sin(\pi \sigma)}{\pi}$. Therefore, for clarity, we define $c^{(2)}_{\Delta,\ell}=-\frac{\sin(\pi \sigma)}{\pi}\tilde{c}^{(2)}_{\Delta,\ell} $. In this normalization, the coefficient $\tilde{c}^{(2)}_{0,0} = 1$. 

Using \eqref{LogToLeft} and \eqref{legcft}, the conformal block decomposition of the CFT Legendre relation becomes
\begin{equation}
\label{CFTLegCBD}
\begin{split}
   \frac{\sin{(\pi \sigma)}}{\pi}  =|1-z|^{-2\sigma(1-\sigma)}\sum_{\Delta,\ell} \frac{\tilde{c}^{(2)}_{\Delta,\ell}}{2} \mathcal{L} \left[ g_{\Delta,\ell}(z,1-\bar{z})\right ]|_{\bar{z} =z} \,
\end{split}
\end{equation}
The above relation gives novel formulae for $1/\pi$ for any $\sigma$ and $z$, among which the fastest, obtained when $\sigma = \frac{1}{2}$ and $z=\bar z=1/2$, takes the following form. 
\begin{equation}
\label{LegCBDexp}
\frac{1}{\pi} 
\approx \frac{a_0+a_1\log 2+a_2\log^2 2}{\sqrt{2}}
\end{equation}
where $a_0, a_1, a_2$ are rational numbers, and we have retained finitely many operators in the conformal block expansion. For instance, retaining the leading 8 operators \cite{foot1} in \eqref{LegCBDexp}.  leads to 6 decimal places accuracy, which is in fact comparable to \eqref{rampi}. 

While novel and interesting in their own right, formulae resulting from \eqref{CFTLegCBD} do not converge fast. This happens because \eqref{logcorCBD} requires $z,\bar{z} \ll 1$ in $g_{\Delta,\ell}(z,\bar{z})$ for fast convergence, while the Legendre relation involves $g_{\Delta,\ell}(z,1-z)$. This is essentially the same problem that Ramanujan faced and resolved by expressing the Legendre relation only in terms of $F_{\sigma}(z)$ in \eqref{final}. One could just expand \eqref{final} in conformal blocks using \eqref{logcorCBD} and get fast-converging series, especially at large $n$.  For example, retaining only the log-identity operator, $n=58$ gives new formulae for $\pi$ with 18 decimal places accuracy (see \eqref{pi18}) while $n=1024$ (which can be worked out using a recursion relation \cite{borwein}) gives 84 decimal places. We give more details in the supplementary material \cite{supp}. However, our goal is to obtain a fast converging basis for the Legendre relation solely via CFT methods, and therefore, also provide a physics explanation for the existence of Ramanujan-like formulae parametrised by $n$. This is what we will discuss next.

 {\bf A new dispersive representation for LCFTs:}
Unlike the full correlator, the discontinuity of the correlator across the logarithmic cut given in \eqref{LogToLeft} only involves $F_{\sigma}(z)$ and $F_{\sigma}(\bar{z})$ and therefore, has a fast-converging block expansion.  This directly points us towards the well-established dispersion relations for CFT correlators, which allow us to reconstruct the full CFT correlator just from its discontinuity. In particular, we consider the so-called stringy/parametric dispersion relation recently proposed and derived in \cite{SS, bcss, stringydisp}, as it naturally leads to a one-parameter family of dispersive representations for CFT correlators. 

\textit{The Stringy Dispersion Relation}: If a function $F(x,y)$ is $x \leftrightarrow y$ symmetric and satisfies $\lim_{x \to \infty}|F(x,y)| \to 0$ for $y \in \mathcal{D}$, then for all $-\lambda \in \mathcal{D}$, \cite{ZahedAS, GSZ, RamanAS, BissiSinha, song, SS, bcss}:
\begin{equation}
 F(x,y) =  \frac{1}{\pi} \int_{\mathcal{C}} d\xi H^{(\lambda)}(\xi,x,y)~F_{disc}^{(x)}\left(\xi, \eta^{(\lambda)}(\xi,x,y)\right)   
\end{equation}
where the integral is over all the branch cuts $F(x,y)$ may have in $x$, and $F_{disc}^{(x)}(x,y)$ is the discontinuity of the function across the branch cuts.  When the cuts are along the real line, it is simply $\displaystyle \lim_{\epsilon \to 0}\frac{F(x+i \epsilon,y)-F(x - i \epsilon,y)}{2i}$, which for real analytic functions equals $Im F(x,y)$. We have also defined:
\begin{equation}
\begin{split}
 H^{(\lambda)}(\xi,x,y) =  & \left(\frac{1}{\xi-x}+ \frac{1}{\xi-y} - \frac{1}{\xi+ \lambda}\right)\,, \\
    \eta^{(\lambda)}(\xi, x,y) ~ = ~ & \frac{(x+\lambda)(y+\lambda)}{\xi+{\lambda}} -\lambda \,.
\end{split}
\end{equation}
The representation is independent of the parameter $\lambda$ --- see \cite{stringydisp} for a neat proof and \cite{supp} for numerical checks. 

Now, let us apply the above dispersion relation to the LCFT correlator. We first define $F_{L}(z,\bar{z}) = \kappa(z,\bar{z})^{-1} G_{L}(z,\bar{z}) =  F_\sigma(z)F_\sigma(1-\bar{z})$. Now, $F_{L}(z,\bar{z})$ is symmetric under $z \leftrightarrow 1 -\bar{z}$ and goes to zero at large $z$ for any $\sigma$ \cite{foot2}. So, it can be expressed using the above dispersion relation as
\begin{equation}
\begin{split}
\label{CorrDisp}
       F_L(z,\bar{z}) = \int_{1}^{\infty}  d\xi &\, H^{(\lambda)}(\xi, z, 1-\bar{z})  \\&F_{log} \left(1-\xi, \eta^{(\lambda)}(\xi,z,1-\bar{z})\right)\,.
\end{split}
\end{equation}
This converges for $Re (\lambda) > -1$. We have used the fact that the discontinuity of $F_{L}(z,\bar{z})$ across the branch cut along $z \in (1, \infty)$ is given by $\pi F_{log}(z,\bar{z}) = \sin(\pi \sigma)F_\sigma(z) F_\sigma(\bar{z})$ using \eqref{LogToLeft}.

Now we can express the Legendre relation solely in terms of the discontinuity of the correlator as
\begin{equation}
\label{LegDisp}
\begin{split}
       \frac{\sin(\pi \sigma)}{\pi} = \int_{1}^{\infty} &   d\xi\, \mathcal{L} \bigg[ H^{(\lambda)}(\xi, z, 1-\bar{z})  \\&F_{log} \left(1-\xi, \eta^{(\lambda)}(\xi,z,1-\bar{z})\right) \bigg]  _{\bar{z} = z}\,.
\end{split}
\end{equation}
Performing this integral analytically is hard in general, but a simple analytical verification is possible in the large $\lambda$ limit, since the integrand can be shown to be a total derivative as follows.
\begin{eqnarray}
\label{LargeLamCSCBD}
    &&\frac{\sin\pi\sigma}{\pi}\mathcal{L}\int_1^\infty d\xi H^{(\infty)}(\xi,z,w_{\bar{z}}) F_\sigma(1-\xi)F_\sigma(z+w_{\bar{z}}-\xi){\bigg|}_{\bar z=z}\nonumber \\&=&-z(1-z)\frac{\sin\pi\sigma}{\pi}\int_1^\infty d\xi \,\partial_\xi \left[H^{(\infty)}(\xi,z,w_{\bar{z}})F_\sigma(1-\xi)^2\right]\,,\nonumber\\
    &=&\frac{\sin\pi\sigma}{\pi}\,.
\end{eqnarray}
where $w_{\bar{z}} = 1- \bar{z}$ and in the last line follows because the entire contribution comes from the lower limit. 

\begin{figure}[H]
	\centering
	\includegraphics[scale=0.5]{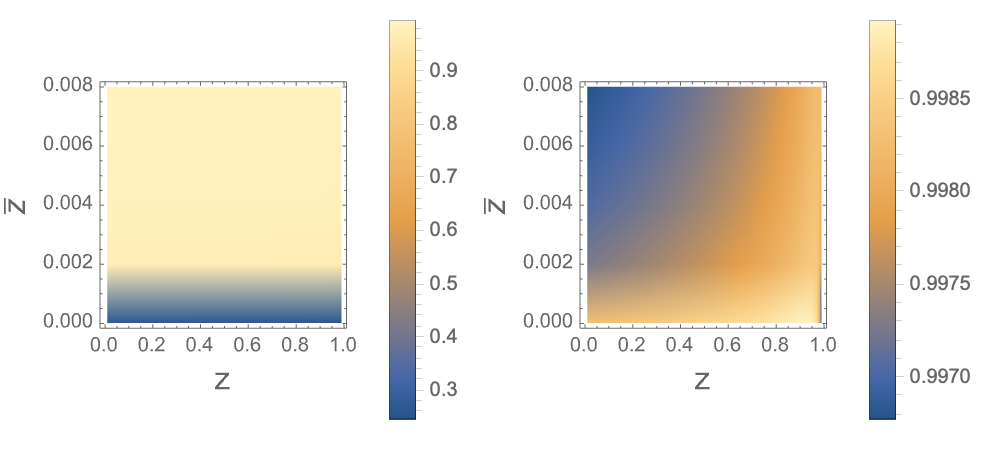}
	\caption{Heat plot showing ratio of sum of blocks ($\Delta=2n+\ell$, $n\leq 2,\ell\leq 4$) to exact answer. Left: CBD, Right: Dispersive representation ($\lambda=500$).}
    \label{heat}
\end{figure}

The next step is to plug the following conformal block decomposition of the discontinuity (see \eqref{logcorCBD}) in \eqref{CorrDisp}. This leads to a new family of bases for the expansion of the LCFT correlator, parametrised by $\lambda$ \cite{foot3}.
\begin{equation}
\label{FlogCBD}
  F_{log} \left(z, \bar{z}\right)  = |1-z|^{-\sigma(1-\sigma)}\sum_{\Delta,\ell} \frac{c^{(2)}_{\Delta,\ell}}{2} g_{\Delta,\ell}(z,1-\bar{z})
\end{equation}
Although the expansion is independent of $\lambda$, convergence improves significantly when $\lambda$ is taken to be large. In fact, as we demonstrate in the heat plot in Fig. \ref{heat}, it is significantly better than the usual conformal block decomposition, especially in the small $\bar{z}$ region.

Finally, plugging \eqref{FlogCBD} in \eqref{LegDisp} leads us to a family of expansions of the Legendre relation in terms of the conformal data, parametrised by $\lambda$.  These expansions are fast-converging when $\lambda$ is taken large. To see this, note that they involve conformal blocks in the form $g_{\Delta,\ell}\left(1-\xi, \eta^{(\lambda)}(\xi,z,1-\bar{z})\right)$. For fast convergence, we need both the arguments of the blocks to be small. In the large $\lambda$ limit, $\eta^{(\lambda)}(\xi,z,1-\bar{z}) \sim 1+z-\bar{z}-\xi$, and since the Legendre relation requires $\bar{z} =z$, both the arguments become $1-\xi$. Now we just need the integral to be dominated by $\xi = 1$, which happens because the kernel $H^{(\lambda)}(\xi, z, 1-\bar{z})$ peaks near $\xi \sim 1$ and more so when $\bar z\sim 0$. Therefore, via purely CFT methods, we have obtained a family of fast-converging expansions of the Legendre relation, just like the derivation of Ramanujan's $1/\pi$ formulae that we reviewed before.

\begin{figure}[H]
    \centering
    \includegraphics[scale=0.5]{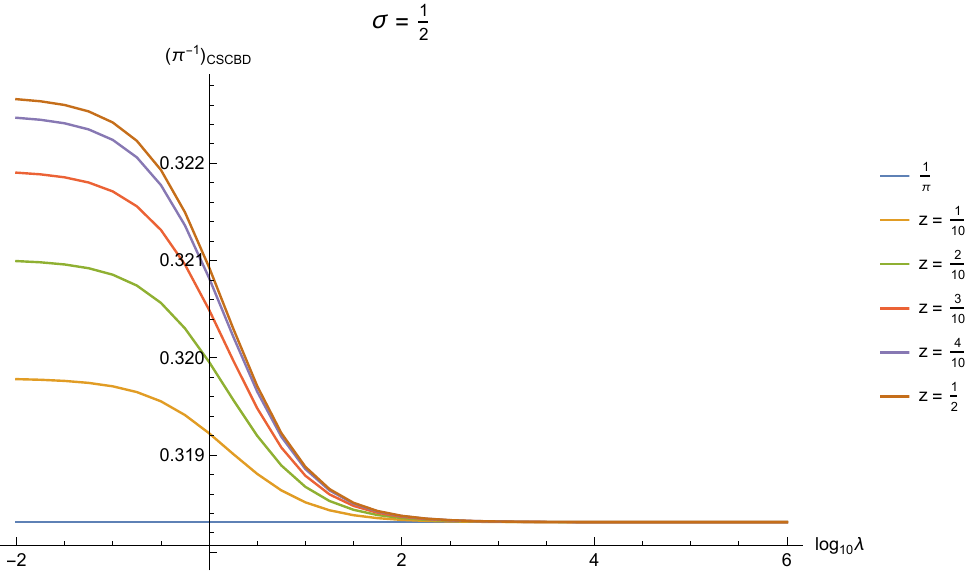}
    \caption{Plot showing the contribution of the log-identity operator in the dispersive block expansion of $1/\pi$ as $\lambda$ is increased for $\sigma =1/2$ and various $z$. We observe that at large $\lambda$, the log-identity operator captures the full Legendre relation, giving $1/\pi$.}
    \label{Pi_Conv_00LargeLambda_1}
\end{figure}

In fact, quite strikingly, the improvement in convergence is so drastic, that in the strict $\lambda \to \infty$ limit, \textit{the entire contribution to the Legendre relation comes solely from the log-identity operator} \cite{foot4}. This can be shown analytically. Plugging the block expansion from \eqref{logcorCBD} in the second line of \eqref{LargeLamCSCBD}, we get
\begin{equation}
\begin{split}
  z(1-z)\int_1^\infty d\xi \,\partial_\xi & \Big[H^{(\infty)}(\xi,z,w_{\bar z}) \Big((1-\xi)^{-2\sigma(1-\sigma)} \\
    & \sum _{\Delta, \ell} \frac{c^{(2)}_{\Delta,\ell}}{2} g_{\Delta,\ell}(1-\xi, 1-\xi)\Big)\Big]\,. 
\end{split}
\end{equation}
This is a total derivative, and only the lower limit $\xi \to 1$ contributes. Since $g_{\Delta,\ell}(1-\xi, 1-\xi)\underset{\xi \to 1}{ \sim}  (1- \xi)^{\Delta +\ell}$, the entire contribution comes from $\Delta = \ell = 0$. We also demonstrate this numerically in Fig. \ref{Pi_Conv_00LargeLambda_1} for $\sigma=1/2$

This remarkable result also enables us to get an approximate (which becomes increasingly accurate as $n$ increases) LCFT handle on the modular equations as discussed in the Appendix. Furthermore, the feature that allowed us to express the discontinuity of the correlator itself in terms of conformal blocks was just the logarithmic nature of the branch cut. Therefore, the above result is likely a general feature of LCFT correlators. We emphasise that it only arises when the correlator is expressed via the stringy dispersion relation and then acted upon by the $\mathcal{L}$ operator --- operations that were motivated by insights from Ramanujan's formulae.

{\bf{Discussion:}}
We have seen how CFT considerations shed new light on Ramanujan's formulae and, in turn, how we are led to new, fast converging expansions of LCFT correlators in terms of the CFT data, using lessons learned from Ramanujan's formulae. It will be interesting to exploit this connection to set up the bootstrap program for LCFTs. In fact, given that the log-identity operator reproduces the full CFT Legendre relation in the large $\lambda$ limit, we expect the LCFT bootstrap to completely fix the full LCFT correlator. 

Additionally, in the Appendix, we show that the dispersive representation with the log-identity operator gives a very good approximation to the modular equations. Do the modular equations and singular values reviewed above carry a deeper physical significance beyond their mathematical roles? We would like to point out that in Saleur's work \cite{saleur}, $G_L$ and $G_R$ can be thought of as configurations of polymers stretching horizontally and vertically. The modular equation can be reinterpreted as the probability of the vertical configurations being a factor of $n$ larger than the horizontal ones. Hence, in the $n\gg 1$ limit, we have a chiral polymer configuration. In the Appendix, we also give a holographic interpretation of the $\sigma=1/2$ Legendre relation in terms of the membrane paradigm. We hope these observations will motivate further research into these remarkable mathematics-physics connections.


\section*{Acknowledgments}
We thank B. Ananthanarayan, Agnese Bissi, Justin David, Debhashis Ghoshal, Rajesh Gopakumar, Srijan Kumar, Rob Myers, Sridip Pal, Prashanth Raman, Arnab Saha, Chaoming Song, Spenta Wadia, Masaaki Yoshida and Xinan Zhou for useful discussions. 
We also thank Gaurav Bhatnagar and Kaushal Verma for raising related questions which spurred this research. AS thanks Perimeter Institute for support and hospitality.
 AS acknowledges support from SERB core grant CRG/2021/000873 and from a Quantum Horizons Alberta chair professorship. 

\section*{Appendix}
{\bf{More about the $c =-2$ LCFT}}: The LCFT we focus on in this work is the well-studied $c = -2$ symplectic fermion model \cite{saleur, symplectic} \cite{foot5}. The model consists of a two-component fermionic field $ \chi^{\alpha}$ whose modes $\chi^{\alpha}_n, n \in \mathbb{Z}$ generate a chiral algebra. The Virasoro algebra with $c = -2$ is contained within this algebra. The algebra has a unique irreducible highest-weight representation, with the highest-weight state being the vacuum state $\Omega$. This representation can be extended to obtain reducible but indecomposable representations, which contain, in addition to the vacuum state $\Omega$, another state $\omega$. The states $\omega$ and $\Omega$ lead to a two-dimensional Jordan block structure for $L_0$.
\begin{equation}
    L_0 \Omega = 0, \quad L_0 \omega = \Omega
\end{equation}
It is this non-diagonalizability of the $L_0$ operator that leads to logarithmic correlation functions in LCFTs. 
 
The symplectic fermion model further has a global $SL(2,\mathbb{C)}$ symmetry which allows us to introduce twisted states by quotienting with respect to the cyclic sub-groups $\mathcal{C}_N$. The twist fields $\mu_\sigma$ are defined by
\begin{equation}
    \chi^{\alpha}(e^{2\pi i} x) \mu_\sigma(0) \sim e^{2\pi i \sigma}  \chi^{\alpha}(x) \mu_\sigma(0) 
\end{equation}
where $\sigma$ is the twist. For the $\mathcal{C}_N$-twisted models, the highest-weight twist states have $\sigma= \frac{k}{N}, k = 1,..,N-1$ and conformal weights $h = -\frac{\sigma(1-\sigma)}{2}$. Their OPE features the logarithmic doublet $(\Omega,\omega)$
\begin{equation}
\label{TwistOPE}
    x^{\sigma(1-\sigma)} \mu(x) \mu(0) \sim \omega + \Omega \log x + ~ . . .
\end{equation}
The $\mathcal{C}_2$-twisted model has been shown to be equivalent to the well-known triplet model \cite{symplectic}, which is an extension of the Virasoro $(1, 2)$ algebra by a triplet of fields $W^a(z), ~a =1,2,3$. In general, all Virasoro $(1, q)$ models have been shown to be extended to triplet algebras and lead to rational LCFTs \cite{kausch, gaberdiel1}. Although we consider the $q = 2, c = -2$ case in this work, the connection to Ramanujan formulae also exists for the other Virasoro $(1, q)$ models. We explain the connection to LCFTs of other central charges in \cite{supp}. 

As is evident from \eqref{TwistOPE}, the correlation functions of twist fields have logarithmic singularities. In particular, the correlator considered in \eqref{LCFTcorr} is the four-point correlator of two twist and two anti-twist fields (with twists $\sigma^* = 1- \sigma$).
\begin{equation}
\label{TwistCorrDef}
\begin{split}
    &G(z,\bar{z})= \langle \mu_{\sigma}(0) \mu_{\sigma^*}(z,\bar{z}) \mu_{\sigma}(1)\mu_{\sigma^*}(\infty)\rangle\,.
\end{split}
\end{equation}
Note that while the twist and anti-twist fields are distinct, they have the same conformal weights. $\sigma = \frac{1}{2}$ is the exception and leads to the correlator of identical scalar primaries with $h = -\frac{1}{8}$.

{ \bf A physical interpretation of the Legendre Relation via holography}:  Consider a scalar $\varphi$ in an asymptotically $AdS_{d+1}$ Schwarzschild black brane background.  
\begin{equation}
\label{AdSBBMetric}
ds^2 = \frac{R^2}{z^2} \left[ -(1 - z^d)dt^2 + \frac{dz^2}{1 - z^d} + d\vec{x}^2 \right]
\end{equation}
where $R$ is the $AdS$ radius and $z$ is the radial bulk coordinate \cite{grinberg}.
We fix the mass of the scalar so that it satisfies the Breitenlohner-Freedman bound, i.e.  $m^2 =  -d^2/(4R^2)$. It turns out that, under these conditions, the bulk-bulk Green's function of the scalar has exactly the structure of an LCFT correlator. More precisely, consider the Fourier modes of the bulk-bulk Green's function in the $\vec{x}, t$ directions, denoted as $\hat{G}(z, z', \omega, \vec{k})$. The zero-mode $\hat{G}(z,z') = \hat{G}(z,z',\omega,\vec{k})\big|_{\omega=\vec{k}=0}$ is especially physically significant and satisfies
\begin{equation}
\label{GreenEq}
    \partial_z \left( \frac{R^{d-1}}{z^{d-1}} (1-z^d) \partial_z G(z, z') \right) - \frac{R^{d+1}}{z^{d+1}} m^2 G(z, z') = \delta(z - z')
\end{equation}
The two linearly independent solutions of the homogeneous part of this differential equation are given by
\begin{equation}
    g_{inf}(z) = z^{1/2}F_{\frac{1}{2}}(z), \quad  g_{hor}(z) = z^{1/2}F_{\frac{1}{2}}(1-z)
\end{equation}
and are called so because $g_{inf}(z)$ is regular at the $AdS$ boundary at $z = 0$ (with a logarithmic singularity at $z =1$), while $g_{hor}(z)$ is regular at the black brane horizon at $z = 1$ (with a logarithmic singularity at $z = 0$). 
From here, $G(z,z')$ can be computed and is given as
\begin{equation}\label{cases}
G(z,z') =
\begin{cases}
 -\dfrac{\pi}{d\,R^{d-1}} g_{inf}(z) g_{hor}(z') , & z' > z, \\
- \dfrac{\pi}{d\,R^{d-1}}g_{hor}(z) g_{inf}(z') & z < z',
\end{cases}
\end{equation}
Thus, we see that the bulk-bulk Green's function has precisely the structure of an LCFT correlator. 

The Legendre relation in this context is simply the Wronskian of the homogeneous part of the differential equation. 
\begin{equation}\label{wrons}
  W(z) = g_{inf}(z) \partial_z g_{hor}(z)-  g_{hor}(z) \partial_z g_{inf}(z) = -\frac{1}{\pi} \frac{1}{1-z}
\end{equation}
One can check that the above equation is exactly the same as the Legendre relation in \eqref{leg} or \eqref{legcft} with $\sigma = 1/2$.

\paragraph{Physical role of the Wronskian and the static susceptibility.}
Together with the Green–function eq.(\ref{GreenEq}), eq.(\ref{wrons}) implies a radially conserved symplectic “flux’’ for the pair $\{g_{\rm inf},g_{\rm hor}\}$. Integrating eq.(\ref{GreenEq}) across $z=z'$ gives the usual jump condition on $\partial_z G$, and substituting the piecewise form \eqref{cases} with the Wronskian identity \eqref{wrons} fixes the overall normalization of $G$ independent of where it is evaluated. In boundary terms, this same conservation makes the \emph{static susceptibility} of the dual operator at $\omega=k=0$ purely IR–determined: expanding the IR–regular bulk solution near the AdS boundary  as $\phi(z)=z^{d/2}\!\big(A\,\log(\mu z)+B+\cdots\big)$ (where $A$ is the source coefficient and $B$ the response), horizon regularity plus the constant Wronskian fix the ratio $B/A$, so $\chi(0)=B/A$ is real and scheme–dependently constant along the radial direction. This is the same “radial Gauss–law’’ logic that underlies horizon formulas for transport (membrane paradigm), specialized here to the scalar static limit: the conserved Wronskian turns horizon regularity directly into the boundary zero–frequency response (see e.g., \cite{sean}).

\paragraph{Order of the modular equation: bulk interpretation.}
In AdS language, an order–$n$ modular equation is most naturally read as passing to the \emph{$n$-fold Euclidean cover} \cite{lewk} of the black–brane thermal circle (the replica “cigar”). Along the modular slice used in \eqref{mod}, this cover renders the two radial bases $g_{\rm inf}$ and $g_{\rm hor}$ \emph{equivalent up to fixed algebraic factors}, and at the singular value $z_0$ they \emph{align}, so the mixed–channel expression collapses to a single channel. Operationally, this is precisely why the step from \eqref{mod} to \eqref{final} is possible and why the resulting series converges parametrically faster: on the $n$-fold cover each term effectively accounts for $n$ windings at once. 

\paragraph{Holographic meaning of the dispersive formula.}
Eq. \eqref{CorrDisp} is a spectral reconstruction of the static bulk–bulk Green function from the
\emph{horizon cut}. The integration variable $\xi\in[1,\infty)$ runs along the branch cut through
$z=1$ of the radial hypergeometric solutions. The $F(1-\xi)$ factor is the
\emph{analytic continuation} $g_{inf}$ across the horizon onto the second Riemann sheet; this is a reflection of the fact that the cut of the UV block is proportional to the IR block. The  kernel factor
propagates this cut (absorptive) data to the bulk points $(z,\bar z)$. 
In this language the horizon cut supplies the spectral density (ingoing condition),
while the conserved flux fixes the static normalisation; together they reproduce the physical
Green function.

The parameter $\lambda$ labels a one–parameter family of dispersion representations of the same
Green function: changing $\lambda$ reshuffles the weight between the two radial channels inside
the integrand but leaves $G$ invariant. Differentiating Eq.\,\eqref{CorrDisp} with respect to $\lambda$ shows
that the variation of the kernel and of its argument combine into a total $\xi$–derivative plus the
radial equation of motion for the hypergeometric block. The bulk term vanishes by the equation
of motion, and the total $\xi$–derivative contributes only boundary terms at the endpoints of the
cut, which vanish. Equivalently, in
holographic language, changing $\lambda$ amounts to adding a homogeneous solution whose
coefficient is fixed to zero by the ingoing (horizon) condition and the Wronskian normalisation.
Thus $G$ is $\lambda$–independent: $\lambda$ is a dispersive “gauge’’ choice rather than new
physics.

{ \bf $1/\pi$ formulas from CFT}: The explicit formula from the log-identity operator contributing to \eqref{final} in the main text for $n =7$ is quite simple:
\begin{equation}
    \frac{1}{\pi} \approx \frac{8\sqrt{7}-11}{8 \sqrt{3 \sqrt{7}+8}}
\end{equation}
giving 5 decimal places. The $n=58$ case gives:
\begin{equation}\label{pi18}
\begin{split}
  \frac{1}{\pi} \approx  &   \frac{1}{2} \Big(13233864246 \sqrt{2}-18715707421\\
  &+ 87 \sqrt{29} \left(28247341 \sqrt{2}-39947352\right) \Big)^{\frac{1}{2}}\,.
\end{split}
\end{equation}
This leads to 18 decimal places agreement. While easy to explain, the formula for $n=1024$, which leads to 84 decimal places for the log-identity operator, is too cumbersome to display. 

{\bf Modular equations from CFT}:
The dispersive representation gives an approximate handle on the modular equations. Using the dispersive representations of $F_\half(z)F_\half(1-\bar z)$ and $F_\half(\bar z)F_\half(1-z)$, retaining only the log-identity operator in the $\lambda\gg 1$ limit, we can make contour plots of the ratio:

\begin{figure}[H]
	\centering
	\includegraphics[scale=0.5]{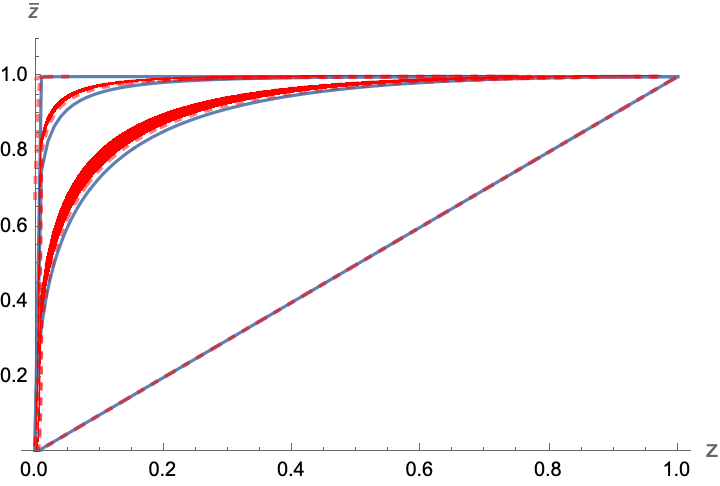}
	\caption{Contour plot of the ratio of $G_L(z,\bar z)/G_R(z,\bar z)$. The blue solid lines and red dashed show the $n=1,2,3,7$ as in fig.\ref{slice}. The thickened red-lines shade the regions between $1.9$ (lower), $2$ (upper) and $2.9$ (lower), $3$ (upper). }\label{modcft}
\end{figure}
As is evident, the log-identity operator on its own provides a very good approximation to the $z=f_n(\bar z)$ slices obtained from the modular equations. 
As $n$ increases, the CFT approximation provides a more accurate solution. For example, the percentage deviations at $z=1/100$ from the known solutions at $n=3,7,23$ are $10\%,0.02\%,10^{-7}\%$ respectively.

\onecolumngrid
\section*{Supplementary material}

\section{More on modular equations}
In the main text, we briefly discussed the modular equations. The simplest example of a modular equation that Ramanujan gives in his notebooks \cite{modnaika} is to start with:
\begin{equation}
F(x)\equiv {}_1F_0(\half;x)=\sum_{n=0}^\infty \frac{(\half)_n}{n!}x^n\,,
\end{equation}
which satisfies
\begin{equation}
F(\frac{2t}{1+t})=(1+t)F(t^2)\,.
\end{equation}
Then $z\equiv 2t/(1+t), \bar z\equiv t^2$ satisfy the modular equation of degree 2:
\begin{equation}
    \bar z(2-z)^2=z^2\,,
\end{equation}
and $(1+t)$ is called the multiplier.
We present a few more nontrivial examples using the notations in the main text which arise from complicated theta function identities \cite{borwein}. 
\begin{eqnarray}
    n&=&2; \quad \bar z=\frac{4 \sqrt{z}}{(1+\sqrt{z})^2}\,,\\
    n&=&3; \quad  (z \bar z)^{1/4}+(1-z)^{1/4}(1-\bar z)^{1/4}=1\,.\\
    n&=&7;\quad
   (z \bar z)^{1/8}+(1-z)^{1/8}(1-\bar z)^{1/8}=1\,.
\end{eqnarray}
Once we are given the equations, we can use
\begin{equation}
    M_n^2(\bar z)=\frac{\bar z(1-\bar z)}{z(1-z)}\frac{d z}{d\bar z}\,,
\end{equation}
to evaluate $M_n(\bar z)$ using the method of implicit differentiation. We note the following explicit results:
\begin{eqnarray}
    \frac{d M_2(\bar z)}{d\bar z}{\bigg |}_{\bar z=1-z_0}&=& -\frac{2+\sqrt{2}}{4}\,,\quad z_0=3-2\sqrt{2}\approx 0.172\,,\\
    \frac{d M_3(\bar z)}{d\bar z}{\bigg |}_{\bar z=1-z_0}&=& -\frac{4}{\sqrt{3}}\,,\quad z_0=\frac{2-\sqrt{3}}{4}\approx 0.0670\,,\\
    \frac{d M_7(\bar z)}{d\bar z}{\bigg |}_{\bar z=1-z_0}&=& -\frac{80}{\sqrt{7}}\,,\quad z_0=\frac{8-3\sqrt{7}}{16}\approx 0.00392\,,
    \end{eqnarray}

We expect that since the conformal block expansion in the main text corresponds to the s-channel OPE, as we increase $n$, will get more and more digits of $1/\pi$ for the same number of operators. In fact, we expect the log-identity operator alone gives the entire contribution at large $n$. Indeed this seems to be the case and we tabulate our findings below (for $\sigma=1/2$ case):

\begin{table}[H]
\centering
\begin{tabular}{|c|c|c|}
\hline
$n$ & 10 ops. & 1 op \\
\hline
1 & 7 & 1\\
\hline
2 & 11 & 2\\
\hline
3 & 15 & 3\\
\hline
7 & 24 & 5\\
\hline
12 & 33 & 7\\
\hline
58 & 78 & 18\\
\hline
64 & 82 & 19\\
\hline
1024 & 344 & 84\\
\hline
\end{tabular}
\caption{Number of decimal places agreement with 10 non-zero operators and just the log-identity operator in the conformal block decomposition. We have added the results for $n=12,58,64,1024$. The $n=64,1024$ follow from a recursion relation given on pages 160,161 in \cite{borwein}. For the $n=1024$ case, $z_0\sim O(10^{-43})$.}
\end{table}

\section{Checks on the dispersion relation}
The dispersive representation used in the main text follows from the analysis in \cite{RamanAS, BissiSinha, SS, bcss}, and the main formula was motivated from string theory \cite{SS, bcss}. There is an independent mathematical proof in \cite{rosengren} for the string theory motivated formulae found in \cite{SS}. It will be nice to see if alternative proofs of our dispersive representations can be found. 
Since the application of the dispersive representation in the CFT context may be unfamiliar, we will first list out a few nontrivial numerical checks, in case the reader wishes to perform similar checks. Explicitly let us use $F(z,w)=F_\half(z)F_\half(w)+F_\half(1-z)F_\half(1-w)$, for which we have:
\begin{equation}\label{disp5}
    F(z,w)  = \frac{1}{\pi} \int_{1}^{\infty}  d\xi\, H^{(\lambda_1)}(\xi, z, w)F_{\frac{1}{2}}(1-\xi)  F_{\frac{1}{2}}( \eta^{(\lambda_1)}(\xi,z,w))-\frac{1}{\pi} \int_{-\infty}^{0}  d\xi\, H^{(\lambda_2)}(\xi, z, w)F_{\frac{1}{2}}(\xi)  F_{\frac{1}{2}}(1- \eta^{(\lambda_2)}(\xi,z,w))
\end{equation}
We will need $Re(\lambda_1)>-1, Re(\lambda_2)<0$ to avoid spurious singularities.
 \begin{enumerate}
\item Choosing $\lambda_2=-\lambda_1$ and dialling $\lambda_1$ from 1 to 100, we have verified (the Mathematica precision we used was AccuracyGoal=50, PrecisionGoal=10, MaxRecursion=50, WorkingPrecision=30)
\begin{equation}
    F(\frac{1}{10},{\frac{3}{10}})=3.2884898786\,,\quad F(\frac{1}{2},\frac{1}{2})=2.7864078594\,,
\end{equation}
which are the expected answers.
\item We chose $\lambda_2=-\lambda_1>0$ to avoid introducing a pole on the integration line. We can also avoid this by choosing $\lambda_2=\lambda_1=p+i$ and dial $p$. This gives rise to the same answers as before. 
\item We have also verified the constancy of the answer with $\lambda_i$ for $z,w$ values that lie above or below the cuts.
 \end{enumerate}

\section{Nested integral representation}

A fascinating consequence of the dispersive representation is the possibility of nesting the integrals. Note that $F_\sigma(x)F_\sigma(y)$'s dispersive representation features again the quadratic $F_\sigma$ inside the integral with different arguments. Hence, we can nest the integrals. For example, the first nesting leads to:
\begin{eqnarray}
    K_\sigma[z,w]\equiv F_\sigma(z)F_\sigma(w)=\frac{\sin^2\pi \sigma}{\pi^2}\int_1^\infty \int_1^\infty d\xi_1\,d\xi_2\, && H^{(\lambda_1)}(\xi_1, z,w) H^{(\lambda_2)}(\xi_2, 1-\xi_1, \eta^{(\lambda_1)}(\xi_1,z,w))\nonumber \\ &\times& K_\sigma[1-\xi_2,\eta^{(\lambda_2)}(\xi_2,1-\xi_1,\eta^{(\lambda_1)}(\xi_1,z,w))]\,.
\end{eqnarray}
This could potentially be a fruitful starting point to investigate novel formulae for $\pi$, since unlike the formula without nesting where the dispersive representation comes with $1/\pi$ in front, this one brings an additional power of $1/\pi$. So applying the Legendre relation will potentially give formulae for $\pi$. This should also be investigated in the context of the bootstrap.

\section{Ramanujan-Orr series}
 
While Ramanujan's original series appears to rely on the Clausen identity, a similar series for $1/\pi$, sometimes called Ramanujan-Orr series follows from the Orr relation \cite{wan}:
\begin{equation}
    F_\sigma(x)F_\sigma(\frac{x}{x-1})={}_4F_3(\sigma,\sigma,1-\sigma,1-\sigma;\frac{1}{2},1,1;-\frac{x^2}{4(1-x)})\,.
\end{equation}
leading to the Ramanujan-Orr series for $1/\pi$ from here reads:
\begin{equation}\label{ramorr}
    \sum_{n=0}^\infty \frac{(\sigma)_n^2 (-\sigma)_n^2}{4^n (\frac{1}{2})_n n!^3}\frac{\sigma^2-3n^2}{\sigma}=\frac{\sin\pi\sigma}{\pi}
\end{equation}.

While the original Ramanujan formulae appear to need $c<0$, the Ramanujan-Orr formula for $\sigma=2/3$ corresponds to $c=0$, which arises in another very well-studied log-CFT corresponding to percolation \cite{cardy, langlands, he}. For the physics behind the Ramanujan-Orr series, we can also have non-unitary CFTs with $c>0$ which corresponds to $2/3\leq \sigma \leq 3/2$ (we will restrict to $\sigma<1$). An example that has been studied in the literature is the case $\sigma=3/4$ which corresponds to $c=1/2$. In \cite{logmin}, this is referred to as the logarithmic Ising model. Another example mentioned in the same reference is $\sigma=4/5$ which corresponds to $c=7/10$ and is referred to as the logarithmic tricritical Ising model.

\section{Connection to LCFTs of other central charges}
Ramanujan $1/\pi$ formulae are not only connected to the triplet algebra at $c = -2$ coming from the Virasoro (1,2) model, but also to rational CFTs of other central charges whose chiral algebras are the triplet algebras arising from Virasoro $(1,q)$ models.  These LCFTs are connected to the minimal models. After identifying $q = 1/\sigma$, they have central charges given by the same formula as for Virasoro $(1,1/\sigma)$ minimal models, i.e. $c = 13-6 (\sigma + \sigma^{-1})$.

For $\sigma = \{1/2, 1/3, 1/4, 1/6\}$, we get $c= \{-2, -7, $ $-25/2, -24\}$ LCFTs. We discussed $c = - 2$ already, $c = -24$ appears in the study of percolation \cite{flohrlohmann}, and $c = - 7$ is briefly discussed in \cite{flohr}, while we have not found any mention of the $c = - 25/2$ theory in any physics context.

To get to the correlators of interest, we consider the degenerate representations in Virasoro $(1,1/\sigma)$ minimal models. The highest-weight states for these representations at level $rs$ have weight $h_{r,s} = \frac{(r-\sigma  s)^2-(1-\sigma )^2}{4 \sigma }$. We will focus on level two states $\Phi_{1,2}$ with weight $h_{1,2} = (3\sigma -2)/4$. Consider now the four-point correlator $\langle \Phi_{r,s}(0)  \Phi_{1,2}(z,\bar z) \Phi_{r,s}(1) \Phi_{1,2}(\infty)\rangle$ in minimal models. The holomorphic part of the correlator satisfies the following second-order BPZ differential equation \cite{BPZ, bigfat} 
\begin{equation}
\begin{split}
   \Big( \frac{z(1-z)}{\sigma} \partial_z^2 + (1-2z) \partial_z -\frac{h_{r,s}}{z(1-z)} \Big)G(z) = 0    \,.
\end{split}
\end{equation}
In general, the solutions to this equation do not have $\log$ singularities. But if we consider $s = r/\sigma$ so that $h_{r,r/\sigma} = -(\sigma-1)^2/4\sigma$, the two solutions become
\begin{equation}
    z(1-z)^{\frac{1-\sigma}{2}} F_{\sigma}(z), \quad z(1-z)^{\frac{1-\sigma}{2}} F_{\sigma}(1-z)\,,
\end{equation}
which have $\log$ singularities. The full correlator is given by 
\begin{equation}
\begin{split}
  \langle \Phi_{r,r/\sigma}(0) \Phi_{1,2}&(z,\bar z) \Phi_{r,r/\sigma}(1) \Phi_{1,2}(\infty)\rangle =  |z(1-z)|^{1-\sigma} F(z,\bar{z})\,, \\
  F(z,\bar{z}) &=F_\sigma(z) F_\sigma(1-\bar z)+F_\sigma(\bar z) F_{\sigma}(1-z)\,.
\end{split}
\end{equation}
Thus, we find a very similar form for the correlator as we had seen for twist $\sigma$ states in the symplectic fermion model. In fact, for $\sigma = 1/2$, we recover the exact same correlator as for weight $h= -\frac{1}{8}$ twist fields in the $\mathcal{C}_2$-twisted model. For other $\sigma$, despite the similar form, these are not related. For example, for $\sigma = 1/4$, while we get the correlator of twist fields with weight $h =-3/32$ in the $\mathcal{C}_4$-twisted model, in this case we get the correlator of fields with weights $h_{1,2} = -5/16$ and $h_{r,r/\sigma} = -9/16$ in a $c =-25/2$ LCFT. The conformal block decompositions also look different. In this case, since the operator weights are different, there is no contribution to the correlator from the identity operator, and consequently, no universality in the Legendre relation in the $\lambda \to \infty$ limit. 


\begin{thebibliography}{99}
\bibitem{ram} S.~Ramanujan, `Modular equations and approximations to $\pi$," 
Quarterly Journal of Mathematics {\bf 45} (1914), 350-372.

\bibitem{pirev} J.~Guillera, `History of the formulas and algorithms for $\pi$", arXiv:0807.0872.

\bibitem{saleur}
H.~Saleur,
`Polymers and percolation in two-dimensions and twisted N=2 supersymmetry,''
Nucl. Phys. B \textbf{382}, 486-531 (1992)
doi:10.1016/0550-3213(92)90657-W
[arXiv:hep-th/9111007 [hep-th]].

\bibitem{gurarie}
V.~Gurarie,
`Logarithmic operators in conformal field theory,''
Nucl. Phys. B \textbf{410}, 535-549 (1993)
doi:10.1016/0550-3213(93)90528-W
[arXiv:hep-th/9303160 [hep-th]].

\bibitem{flohr}
M.~Flohr,
`Bits and pieces in logarithmic conformal field theory,''
Int. J. Mod. Phys. A \textbf{18}, 4497-4592 (2003)
doi:10.1142/S0217751X03016859
[arXiv:hep-th/0111228 [hep-th]].

\bibitem{cardy}
J.~L.~Cardy,
`Critical percolation in finite geometries,''
J. Phys. A \textbf{25}, L201-L206 (1992)
doi:10.1088/0305-4470/25/4/009
[arXiv:hep-th/9111026 [hep-th]].

\bibitem{flohrlohmann}
M.~A.~I.~Flohr and A.~Muller-Lohmann,
`Conformal field theory properties of two-dimensional percolation,''
J. Stat. Mech. \textbf{0512}, P12004 (2005)
doi:10.1088/1742-5468/2005/12/P12004
[arXiv:hep-th/0507211 [hep-th]].

\bibitem{langlands}
R.~Langlands, P.~Pouliot and Y.~Saint-Aubin,
`Conformal invariance in two-dimensional percolation,''
Bull. Am. Math. Soc. \textbf{30}, 1-61 (1994)
doi:10.1090/S0273-0979-1994-00456-2
[arXiv:math/9401222 [math-ph]].

\bibitem{nayak}
V.~Gurarie, M.~Flohr and C.~Nayak,
`The Haldane-Rezayi quantum Hall state and conformal field theory,''
Nucl. Phys. B \textbf{498}, 513-538 (1997)
doi:10.1016/S0550-3213(97)00351-9
[arXiv:cond-mat/9701212 [cond-mat]].

\bibitem{bissiCCFT}
A.~Bissi, L.~Donnay and B.~Valsesia,
``Logarithmic doublets in CCFT,''
JHEP \textbf{12} (2024), 031
doi:10.1007/JHEP12(2024)031
[arXiv:2407.17123 [hep-th]].

\bibitem{grinberg}
M.~Grinberg and J.~Maldacena,
`Proper time to the black hole singularity from thermal one-point functions,''
JHEP \textbf{03}, 131 (2021)
doi:10.1007/JHEP03(2021)131
[arXiv:2011.01004 [hep-th]].

\bibitem{guillera} J.~Guillera, `A method for proving Ramanujan series for $1/\pi$", arXiv:1807.07394.

\bibitem{wolframram} F.~Bhat, A.~Sinha, Mathematica notebook for this paper, Wolfram Community, STAFF PICKS, April 2, 2025, https://community.wolfram.com/groups/-/m/t/3438221

\bibitem{gaberdiel1}
M.~R.~Gaberdiel and H.~G.~Kausch,
`A Rational logarithmic conformal field theory,''
Phys. Lett. B \textbf{386}, 131-137 (1996)
doi:10.1016/0370-2693(96)00949-5
[arXiv:hep-th/9606050 [hep-th]].

\bibitem{gaberdiel2}
M.~R.~Gaberdiel and H.~G.~Kausch,
`A Local logarithmic conformal field theory,''
Nucl. Phys. B \textbf{538}, 631-658 (1999)
doi:10.1016/S0550-3213(98)00701-9
[arXiv:hep-th/9807091 [hep-th]].

\bibitem{gaberdiel3}
M.~R.~Gaberdiel,
`An Algebraic approach to logarithmic conformal field theory,''
Int. J. Mod. Phys. A \textbf{18}, 4593-4638 (2003)
doi:10.1142/S0217751X03016860
[arXiv:hep-th/0111260 [hep-th]].

\bibitem{curious}
H.~G.~Kausch,
`Curiosities at c = -2,''
[arXiv:hep-th/9510149 [hep-th]].

\bibitem{supp} See Supplemental Material for more details, which includes \cite{modnaika, rosengren, wan, he, logmin, BPZ}.

\bibitem{stringydisp}
F.~Bhat, A.~P.~Saha and A.~Sinha,
``A stringy dispersion relation for field theory,''
[arXiv:2506.03862 [hep-th]].

\bibitem{borwein}
J.~Borwein and P.~ J.~ Borwein, `Pi and the AGM: A Study in Analytic Number Theory and Computational Complexity", Canad. Math. Soc. Series Monographs Advanced Texts, (John Wiley, New York, 1987).

\bibitem{abcd}
M.~Hogervorst, M.~Paulos and A.~Vichi,
`The ABC (in any D) of Logarithmic CFT,''
JHEP \textbf{10}, 201 (2017)
doi:10.1007/JHEP10(2017)201
[arXiv:1605.03959 [hep-th]].

\bibitem{foot1}{Explicitly $a_0=\frac{21825486901}{2147483648}, a_1= -\frac{66594751779}{2147483648}, a_2=\frac{52661096730}{2147483648}$.}

\bibitem{SS}
A.~P.~Saha and A.~Sinha,
`Field Theory Expansions of String Theory Amplitudes,''
Phys. Rev. Lett. \textbf{132}, no.22, 221601 (2024)
doi:10.1103/PhysRevLett.132.221601
[arXiv:2401.05733 [hep-th]].

\bibitem{bcss}
F.~Bhat, D.~Chowdhury, A.~P.~Saha and A.~Sinha,
`Bootstrapping string models with entanglement minimization and machine learning,''
Phys. Rev. D \textbf{111}, no.6, 066013 (2025)
doi:10.1103/PhysRevD.111.066013
[arXiv:2409.18259 [hep-th]].

\bibitem{ZahedAS}
A.~Sinha and A.~Zahed,
`Crossing Symmetric Dispersion Relations in Quantum Field Theories,''
Phys. Rev. Lett. \textbf{126}, no.18, 181601 (2021)
doi:10.1103/PhysRevLett.126.181601
[arXiv:2012.04877 [hep-th]].

\bibitem{GSZ}
R.~Gopakumar, A.~Sinha and A.~Zahed,
`Crossing Symmetric Dispersion Relations for Mellin Amplitudes,''
Phys. Rev. Lett. \textbf{126}, no.21, 211602 (2021)
doi:10.1103/PhysRevLett.126.211602
[arXiv:2101.09017 [hep-th]].

\bibitem{RamanAS}
P.~Raman and A.~Sinha,
`QFT, EFT and GFT,''
JHEP \textbf{12}, 203 (2021)
doi:10.1007/JHEP12(2021)203
[arXiv:2107.06559 [hep-th]].

\bibitem{BissiSinha}
A.~Bissi and A.~Sinha,
`Positivity, low twist dominance and CSDR for CFTs,''
SciPost Phys. \textbf{14}, no.4, 083 (2023)
doi:10.21468/SciPostPhys.14.4.083
[arXiv:2209.03978 [hep-th]].

\bibitem{song}
C.~Song,
`Crossing-Symmetric Dispersion Relations without Spurious Singularities,''
Phys. Rev. Lett. \textbf{131}, no.16, 161602 (2023)
doi:10.1103/PhysRevLett.131.161602
[arXiv:2305.03669 [hep-th]].

\bibitem{symplectic}
H.~G.~Kausch,
`Symplectic fermions,''
Nucl. Phys. B \textbf{583}, 513-541 (2000)
doi:10.1016/S0550-3213(00)00295-9
[arXiv:hep-th/0003029 [hep-th]].

\bibitem{gaiotto}
D.~Gaiotto and L.~Rastelli,
`A Paradigm of open / closed duality: Liouville D-branes and the Kontsevich model,''
JHEP \textbf{07}, 053 (2005)
doi:10.1088/1126-6708/2005/07/053
[arXiv:hep-th/0312196 [hep-th]].

\bibitem{kausch}
H.~G.~Kausch,
`Extended conformal algebras generated by a multiplet of primary fields,''
Phys. Lett. B \textbf{259}, 448-455 (1991)
doi:10.1016/0370-2693(91)91655-F

\bibitem{sean}
S.~A.~Hartnoll,
``Lectures on holographic methods for condensed matter physics,''
Class. Quant. Grav. \textbf{26}, 224002 (2009)
doi:10.1088/0264-9381/26/22/224002
[arXiv:0903.3246 [hep-th]].

\bibitem{lewk}
A.~Lewkowycz and J.~Maldacena,
``Generalized gravitational entropy,''
JHEP \textbf{08}, 090 (2013)
doi:10.1007/JHEP08(2013)090
[arXiv:1304.4926 [hep-th]].

\bibitem{modnaika} M.~S.~Mahadeva~Naika,`Modular equations in the spirit of Ramanujan," talk in IIIT, Bangalore, 2012.

\bibitem{rosengren}
H.~Rosengren,
``String theory amplitudes and partial fractions,''
Ramanujan J. \textbf{67}, no.2, 26 (2025)
doi:10.1007/s11139-025-01080-z
[arXiv:2409.06658 [math.CA]].

\bibitem{wan} J.~G.~Wan, `Series for $1/\pi$ using Legendre's relation," arXiv:1302.5984.

\bibitem{he}
Y.~He,
`Logarithmic operators in $c=0$ bulk CFTs,''
[arXiv:2411.18696 [hep-th]].

\bibitem{logmin}
P.~A.~Pearce, J.~Rasmussen and J.~B.~Zuber,
`Logarithmic minimal models,''
J. Stat. Mech. \textbf{0611}, P11017 (2006)
doi:10.1088/1742-5468/2006/11/P11017
[arXiv:hep-th/0607232 [hep-th]].

\bibitem{BPZ}
A.~A.~Belavin, A.~M.~Polyakov and A.~B.~Zamolodchikov,
`Infinite Conformal Symmetry in Two-Dimensional Quantum Field Theory,''
Nucl. Phys. B \textbf{241}, 333-380 (1984)
doi:10.1016/0550-3213(84)90052-X

\bibitem{bigfat}
P.~Di Francesco, P.~Mathieu and D.~Senechal,
`Conformal Field Theory,''
Springer-Verlag, 1997.

\bibitem{pap1}
A.~L.~Fitzpatrick, J.~Kaplan, D.~Poland and D.~Simmons-Duffin,
`The Analytic Bootstrap and AdS Superhorizon Locality,''
JHEP \textbf{12}, 004 (2013)
doi:10.1007/JHEP12(2013)004
[arXiv:1212.3616 [hep-th]].

\bibitem{pap2}
Z.~Komargodski and A.~Zhiboedov,
`Convexity and Liberation at Large Spin,''
JHEP \textbf{11}, 140 (2013)
doi:10.1007/JHEP11(2013)140
[arXiv:1212.4103 [hep-th]].

\bibitem{sridip}
S.~Pal, J.~Qiao and S.~Rychkov,
`Twist Accumulation in Conformal Field Theory: A Rigorous Approach to the Lightcone Bootstrap,''
Commun. Math. Phys. \textbf{402}, no.3, 2169-2214 (2023)
doi:10.1007/s00220-023-04767-w
[arXiv:2212.04893 [hep-th]].

\bibitem{rev}
A.~Bissi, A.~Sinha and X.~Zhou,
`Selected topics in analytic conformal bootstrap: A guided journey,''
Phys. Rept. \textbf{991}, 1-89 (2022)
doi:10.1016/j.physrep.2022.09.004
[arXiv:2202.08475 [hep-th]].

\bibitem{alday}
L.~F.~Alday,
`Large Spin Perturbation Theory for Conformal Field Theories,''
Phys. Rev. Lett. \textbf{119}, no.11, 111601 (2017)
doi:10.1103/PhysRevLett.119.111601
[arXiv:1611.01500 [hep-th]].

\bibitem{balt}
B.~C.~van Rees,
`Theorems for the Lightcone Bootstrap,''
[arXiv:2412.06907 [hep-th]].

\bibitem{foot2} {More precisely, it behaves as $z^{-\sigma}$ for $0 \le \sigma \le 1$ and $z^{-1/2}\log(z)$ when $\sigma = 1/2$}

\bibitem{foot3} {The validity of this can be checked numerically following e.g. the analysis in \cite{BissiSinha}}.

\bibitem{foot4}  {The dispersive representation does not appear useful for getting series representations for $1/\pi$ as there is a $1/\pi$ explicitly appearing outside the integral. However, we can reuse the dispersion relation to replace the $F_\sigma F_\sigma$ factors in the integral, which could be used for such a purpose. This idea is briefly discussed in the supplementary material.}

\bibitem{foot5} {This theory also plays a role in 2d topological gravity \cite{gaiotto}.}

\end{thebibliography}
\end{document}